\documentclass[sigconf]{acmart}
\usepackage{booktabs} % For formal tables

\usepackage{caption}
\usepackage{subcaption, multirow}
\usepackage{comment}
\usepackage{float}
\usepackage{xcolor}

\usepackage[ruled]{algorithm2e} % For algorithms

\usepackage{enumitem}
\usepackage{amsmath}

\usepackage{color}
\usepackage{soul}

\usepackage{mwe}    % loads »blindtext« and »graphicx«
\usepackage{subcaption}

\definecolor{plblue}{rgb}{0.13,0.13,0.7}
\definecolor{pdgrey}{rgb}{0.75, 0.75, 0.75}
\definecolor{pblue}{rgb}{0.13,0.13,1}
\definecolor{pgreen}{rgb}{0,0.5,0}
\definecolor{pred}{rgb}{0.9,0,0}
\definecolor{pgrey}{rgb}{0.46,0.45,0.48}
\definecolor{light-gray}{gray}{0.88}
\definecolor{light-green}{HTML}{90EE90}
\definecolor{light-orange}{HTML}{FED8B1}

\usepackage{listings}
\newfloat{lstfloat}{htbp}{lop}
\floatname{lstfloat}{Listing}
\newlist{mylist}{enumerate}{1}
\setlist[mylist]{label*=(RQ\arabic*),ref=RQ\arabic*}
\makeatletter
\newcommand\myitem[1][]{%
  \if\relax\detokenize{#1}\relax
    \item\relax
  \else
    \protected@edef\@currentlabel{RQ#1}%
    \item[(RQ#1)]
  \fi}
\makeatother

\setcopyright{rightsretained}
\copyrightyear{2021} 
\acmYear{2021} 
\setcopyright{acmcopyright}
\acmConference[SIGCSE'21]{52nd ACM Technical Symposium on Computer Science Education}{March 17--March 20, 2021}{Toronto, Canada}
\acmPrice{15.00}
\acmDOI{10.1145/3287324.3287483}
\acmISBN{978-1-4503-5890-3/19/02}

\begin{document}
\title[Unique Exams]{Unique Exams: Designing assessments for integrity and fairness}
% note:
% \author{Anonymous Authors}
% \affiliation{%
% \institution{Institutions}
% }

% \begin{comment}
\author{Gili Rusak}
\affiliation{%
  \institution{Stanford University}
}
\email{gili@stanford.edu}

\author{Lisa Yan}
\affiliation{%
  \institution{Stanford University}
}
\email{yanlisa@stanford.edu}
% \end{comment}

% \renewcommand{\shortauthors}{L. Yan et al.}

\begin{abstract}
Educators have faced new challenges in effective course assessment during the recent, unprecedented shift to remote online learning during the COVID-19 pandemic. In place of typical proctored, timed exams, instructors must now rethink their methodology for assessing course-level learning goals. Are exams appropriate---or even feasible---in this new online, open-internet learning environment? In this experience paper, we discuss the unique exams framework: our framework for upholding exam integrity and student privacy. In our Probability for Computer Scientists Course at an R1 University, we developed autogenerated, unique exams where each student had the same four problem skeletons with unique numeric variations per problem. Without changing the process of the traditional exam, \textit{unique} exams provide a layer of security for both students and instructors about exam reliability for any classroom environment---in-person or online. In addition to sharing our experience designing unique exams, we also present a simple end-to-end tool and example question templates for different CS subjects that other instructors can adapt to their own courses.

\end{abstract}

\begin{CCSXML}
<ccs2012>
<concept>
<concept_id>10003456.10003457.10003527.10003531.10003533.10011595</concept_id>
<concept_desc>Social and professional topics~CS1</concept_desc>
<concept_significance>500</concept_significance>
</concept>
<concept>
<concept_id>10003456.10003457.10003527.10003540</concept_id>
<concept_desc>Social and professional topics~Student assessment</concept_desc>
<concept_significance>500</concept_significance>
</concept>
</ccs2012>
\end{CCSXML}

\ccsdesc[500]{Social and professional topics~CS1}
\ccsdesc[500]{Social and professional topics~Student assessment}

\keywords{Assessment; pedagogy; grading; large classes; undergraduate instruction; rubrics}

\maketitle

\section{Introduction}
One of the most pressing challenges in education during the COVID-19 pandemic has been how to transform fully in-person courses to online, remote ones~\cite{reich2020whatslostcovid}. Students are in highly variable home environments for schoolwork, and they feel more isolated and distanced from both their teachers and peers. Hastily adapted assessments could mistakenly assess student access to a conducive learning environment, instead of student knowledge. Students and instructors alike are deeply concerned about the surge in unpermitted assignment and exam aid on public online platforms for homework and code discussion~\cite{fern2020BUInvestigatingWhether,newton2020AnotherProblemShifting}. These factors, among many others, have created significant barriers in adapting assessments to the new remote classroom.

Exams are among the most difficult classroom experiences to adopt. In many traditional classrooms, students complete paper copies of the same exam in the same 2--3 hour duration, in the presence of exam proctors. There are common variations around this experience (e.g., open-/closed-/limited-notes or versioned exams), but, by and large, the exam experience is a standardized one of limited resource---a relative measurement of individual student learning. By contrast, online exams provide limited proctoring options and are perceived to encourage unpermitted excessive collaboration~\cite{nyt_cheating_opinion}. Nevertheless, exams can still provide a valuable context for students to practice retrieval learning and self-assess progress towards their own learning goals~\cite{karpicke2011retrieval,guskey2003HowClassroomAssessments}.

In this project, we propose the concept of unique exams: an end-to-end framework that takes an instructor-designed exam skeleton as input and outputs for every student a personalized, unique, randomized exam with corresponding answer key (Figure \ref{fig:pipeline}). Our work has two defining features. First, the single question skeleton produces a consistent exam experience for each student despite small unique numerical or lexical perturbations. Second, the deterministic mapping of student to unique exam-solution pair is designed to facilitate exam question clarifications, minimize additional grading workload, and preserve exam integrity by deterring public internet uploads (PDF or text-only) of the exam content.

Our exam framework is designed to minimize disruptions to the exam experience for students, instructors, and graders. In this paper, we share our experiences with deploying unique exams in a CS probability course at an R1 University in Spring 2020, and we describe how this framework can be applied to any course or major. While we built our framework as a reaction to education during the global pandemic, we believe there is ample reason for continuing use once we return to in-person instruction.
As part of this work, we provide an open-source framework\footnote{Codebase will be released in the camera-ready submission.} and suggest best practices for providing equitable testing environments, where students feel that assessments will fairly and properly assess their understanding of the material.

\section{Background and Related Work}\label{sec:related}
Transitioning university learning from in-person to remote requires rethinking of many academic institutions, many of which are centered around the shared physical space of campus. Instructors need to design classroom communication and discourse in the absence of face-to-face interactions~\cite{hollan1992BeingThere,olson2000DistanceMatters,stylianos2009formative}, and students may have largely different access to technology and resources. Furthermore, academic norms like classroom attendance and assessments are no longer feasible nor enforceable in the same ways they were before.

The in-person exam is among the more challenging components to reproduce in a remote classroom. The hurdle to administering exams is not technical---after all, PDFs can easily replace print---but rather logistical and consequently, pedagogical: Many exams are designed to be individually completed, and consequently, exams are high-stakes assessments that comprise a large portion of a student's grade. While in-person exam proctors can discourage violations of academic integrity, remote proctoring is often infeasible or inhibits student privacy~ \cite{ExamMonitorBCM,ExamMonitor,steinbaugh2020ProctorUThreatensUC}. Remote exams also expand the possibilities for unpermitted behavior---not only is there the risk of peer-based collaboration, but the integrity of the entire exam can be compromised through broadcasted, \textit{online} posting of exam problems on public expert crowd-sourcing platforms~\cite{Chegg,StackOverflow,fern2020BUInvestigatingWhether,newton2020AnotherProblemShifting}.

Despite these drawbacks, in our work we explore a feasible version of the remote, open-ended exam because institutionally, we must often differentiate between students on the basis of a letter grade. More practically, our remote classrooms are temporary, and we should try to reuse many components of the in-person experience where pedagogically valuable. Massive Open Online Courses (MOOCs) have circumvented the open-ended exam format for other modes of formative feedback (e.g., peer feedback, automated tutors, multiple-choice quizzes)~\cite{suen2014PeerAssessmentMassive}, but these assessments fall short in the depth of technical knowledge and feedback offered by exams. As a result, online exam integrity has been an active area of research with the increasing popularity of online education ~\cite{jung2009enhanced,krsak2007curbing}.

There are several tools and frameworks designed to deter unpermitted collaboration. Automatic plagiarism detection tools are widely used to detect student assignment similarity with online or peer solutions~\cite{batane2010turning, whale1988plague, yan2018tmoss}. However, these tools work better with long-form coding assessments and cannot mitigate the risks in short-answer exam formats like those in math and theoretical CS, where answers are procedurally and semantically similar. PrairieLearn is a computerized testing framework, where students receive several questions randomly selected from a question bank~\cite{chen2018HowMuchRandomization,zilles2015computerizedtesting} and has been used successfully to deter unpermitted collaboration. Nevertheless, some groups of students may perceive unfairness due to versioned exams having different questions~\cite{emeka2020StudentPerceptionsFairness}. Logistically, large question banks may be inaccessible or difficult to maintain if a course warrants new exams per term.

Inspired by assignments like the Binary Bomb assessment~\cite{parlante2012NiftyAssignments}, our unique exam framework creates distinct exams from a single exam skeleton to mitigate the risk of broadcasted, online exam posting when administering exams remotely. Because our framework is a small adjustment to the exam \textit{content}---and not the exam \textit{medium}---it can be easily adapted to both paper exams and software/IDE exams~\cite{piech2018BlueBookComputerizedReplacement,corley2020PaperIDEImpact}.

\section{Unique Exams}\label{sec:method}

\begin{figure}[tbp]
    \centering
    \includegraphics[width=\linewidth]{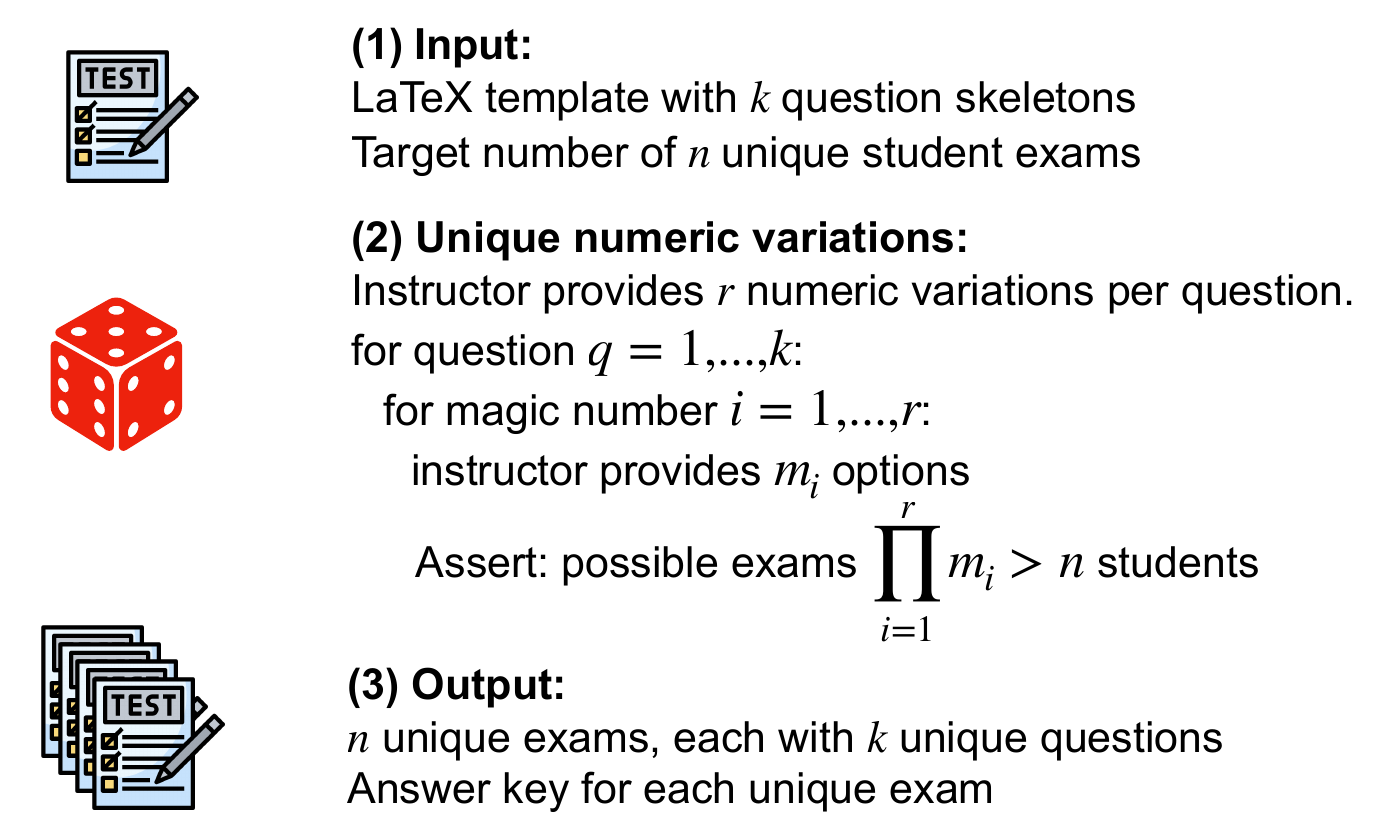}
    \caption{Unique exams allow instructors to provide each student their own, distinct exam. Icons are from \cite{exam_icon}.}
    \label{fig:pipeline}
\vspace{-2ex}
\end{figure}
In this section, we describe the pipeline for our unique exam framework. Because our unique exams were used in a sophomore-level CS probability course (to be described in Section \ref{sec:results}), many of the example problems we describe are computational in nature; however, this pipeline can be generalized to many mathematics and CS exams.
 
\subsection{Overall design} % originally Design Considerations
We had two primary goals in developing our unique exam framework. The first was to uphold the pedagogical goal of exams: a formative assessment where students receive useful feedback on their learning and understanding. All students should have a comparable experience on the exam, in terms of both the course content tested and the medium by which the exam is offered. Second, students should perceive the exam as fair and respectful. The exam experience should be designed to deter students from seeking unpermitted aid without compromising student privacy (e.g., with camera or browser tracking) or being prohibitively expensive. It was not as important for us to design student exam performance as a strong signal for course grades; nevertheless, we also sought a solution that would not incur a significantly more expensive grading process.

After weighing these factors, we developed the unique exam framework, a simple solution that largely leaves the traditional exam structure unaltered, but takes non-invasive measures to encourage honest test-taking behavior culture. In our framework, we design a unique exam \textit{per exam question} and \textit{per student} based on an instructor-develop question skeleton. The identical source skeleton establishes a uniform exam experience---content-wise---because the problem-solving approach to each question is identical, barring numerical or semantic variations. The uniqueness per exam question holds students responsible to adhering to institutional honor code policy. If any subpart of the exam is posted online, instructors can immediately link the online post to a particular student in the classroom.

While this approach is more intensive than simpler solutions (e.g., exam watermarking or versioned exams), our approach can better detect excessive collaboration both when students upload PDF screenshots of exams \textit{and} when they post exam text. Furthermore, since unique exams are a modification on the part of exam content, they can be implemented in conjunction with any exam medium: digital, exam software, printed, or otherwise.

% \gili{Figure 2 should be here.}
\begin{figure*}[ht]
      \centering
      \sethlcolor{light-green}
    \subcaptionbox{Computational problem skeleton.\label{fig:math_ex_skeleton}}
    {
    \colorbox{light-gray}{
    \parbox{3in}{
    \textbf{Problem 1.}
    $X$ and $Y$ are jointly continuous random variables with the following joint PDF:\\
    
    \[ f_{X,Y}(x,y) = c(\text{\hl{$v_1$}} x^2 + \text{\hl{$v_2$}} y) \qquad 0 \leq x\leq \text{\hl{$v_3$}} \text{ and } 0 \leq y \leq \text{\hl{$v_4$}}\]\\
    \sethlcolor{light-orange}
    Note that $f_{X,Y}(x,y)$ is a valid PDF if the constant \hl{$c = v_5$}.\\ 
    
    What is $E[Y]$? Provide a numeric answer (fractions are fine).}
    }
    }
    \subcaptionbox{Computational problem conditions.\label{fig:math_ex_conditions}}
    {
    \begin{tabular}{|c|c|}
     \hline
      Parameter & Type\\
        \hline
      $v_1= c_x = [2,3,4]$ & Independent Parameter \\
      $v_2 = c_y= [2,3,4]$ & Independent Parameter\\
      $v_3 = x_b = [1,2,3]$ &  Independent Parameter\\
      $v_4 = y_b= [1,2,3]$ & Independent Parameter\\
      $v_5 = c = \frac{6}{2v_1(v_3)^3 v_4 + 3v_2 (v_4^2)v_3}$ & Dependent Parameter\\
      solution $E[Y] = v_5\Big(\frac{v_1 (v_3^3)(v_4^2)}{6} + \frac{v_2 v_3 (v_4^3)}{3}\Big)$ & Solution\\
    \hline
    \end{tabular}
    }
    \sethlcolor{light-green}
    \subcaptionbox{Program tracing problem skeleton.\label{fig:code_ex_skeleton}}{
    \colorbox{light-gray}{
    \parbox{3.1in}{
    \textbf{Problem 2.}
    
    \texttt{unsigned char mystery(unsigned char $n$) \{}\\
    \hspace*{0.5ex}\texttt{$\qquad$   $n$ |= $n >> $ \hl{$v_1$};} \\
    \hspace*{0.5ex}\texttt{$\qquad$    $n$ |= $n >> $ \hl{$v_2$};} \\
    \hspace*{0.5ex}\texttt{$\qquad$  $n$ |= $n >>$ \hl{$v_3$};} \\
    \hspace*{0.5ex}\texttt{$\qquad$   $n++$;} \\
    \hspace*{0.5ex}\texttt{$\qquad$ return ($n >> 1$);} \\
    \texttt{\}}
    
    What does the following code print out?

    \hspace*{0.5ex}\texttt{printf("\%u", mystery(\hl{$v_4$})); }
    
    %// \%u prints the integer for an unsigned char
                
    }
    }
    }
    \subcaptionbox{Program tracing problem conditions.\label{fig:code_ex_conditions}}{
    \begin{tabular}{|c|c|}
     \hline
        Parameter  & Type\\
        \hline
       $v_1= [1, 2, 3, 4]$ & Independent Parameter \\
       $v_2=[1, 2, 3, 4]$ & Independent Parameter\\
       $v_3=[1, 2, 3, 4]$ & Independent Parameter\\
       $v_4 = [88, 150]$ & Independent Parameter \\
       solutions are generated programatically & Solution\\
    \hline
    \end{tabular}
    }
    \caption{Example problems with unique values.}
    \label{example_problems}
\end{figure*}

\subsection{Pipeline}
Our system includes several features to provide a simple and easy-to-use framework for instructors to adopt. The pipeline is described in Figure \ref{fig:pipeline}.

The \textit{input} to this framework is a LaTeX template of $k$ problem-solution pairs, where each problem $q$ has $r$ \textit{parameters}. The $i$-th parameter can take on one of a on a set of $m_i$ values and is one of four types:
\begin{enumerate}
    \item Independent parameters: constants,
    \item Dependent parameters: values in the problem that depend on independent parameters,
    \item Conditional parameters: assertions to ensure that the numeric value combinations make sense in the context of the problem.
    \item Solutions: algebraic computation to determine the solution to the unique exam. 
\end{enumerate}

The number of unique problems generated in this process is $\prod_{i=1}^r m_i$, which the instructor must ensure is larger than the $n$ student exams needed. The instructor therefore has several responsibilities: write $k$ problems (and solutions), specify the valid $m_i$ parameters for each parameter $i$ in each problem $q$, and specify any constraints via conditional parameters that would remove logically impossible parameter combinations from the set of unique questions.

The \textit{output} of this framework is: $n$ unique exams with a unique answer key per exam. These exams are generated by taking one unique instantiation of each of the $k$ original problem skeletons. This level of uniqueness is on the granularity of \textit{problems}; in other words, instructors can identify online versions of the exam from just just a portion of the exam allows the teaching staff to identify The batch generation of unique answer keys help the teaching team grading process to proceed similarly to how it is handled in traditional exam settings.

\subsection{Examples}
We provide several examples of problems that pair well with the unique exams framework. Here we show one computational mathematics problem and one code tracing problem skeleton. In the next section, we discuss an end-to-end example of a problem we used in our exam.

\subsubsection{Example 1: Computational mathematics problem} \hfill
\vspace*{1ex}

\textbf{Problem Skeleton.}
Figure \ref{fig:math_ex_skeleton} is a simplified problem skeleton for a mathematics problem used in our case study (Section \ref{sec:results}). The problem contains four independent parameters, $v_1, v_2, v_3, v_4$, and one dependent parameter, $v_5$.

Given the conditions detailed in Figure \ref{fig:math_ex_conditions}, this problem generates $3\cdot 3\cdot 3\cdot 3 = 81$ unique problems. Note that in our actual exam,  we had more choices per independent parameter, and subsequent subsequent subparts of this question included additional independent parameters to generate unique exams for several hundred students.

\textbf{Conditions.}
Figure \ref{fig:code_ex_skeleton} describe the different conditions for developing a set of $k$ unique exams. Here, we specify the rules and options for both the independent and dependent parameters.

\textbf{Solution and Grading.} Grading follows similarly to grading uniform exams. Using the formula described in Figure \ref{fig:code_ex_conditions}, the instructor can generate a detailed solution guide for the teaching team per unique student exam. The teaching team can compare student solutions to the unique answer key when grading with all appropriate steps containing relevant numbers as normal.

\subsubsection{Example 2: Code tracing problem} \hfill
\vspace*{1ex}

\textbf{Problem skeleton.}
This problem (Figure \ref{fig:code_ex_skeleton}) was presented in an Introduction to Computer Systems exam, which we did not deploy in an unique exam setting. \footnote{The authors thank Prof. Chris Gregg for this problem.} 

\textbf{Conditions.}
The conditions are detailed in Figure \ref{fig:code_ex_conditions}.
The following Problem Skeleton and Conditions form $4\cdot 4 \cdot 4 \cdot 4 \cdot 2 = 128$ unique exams.

\textbf{Solution and Grading.}
Grading would proceed similarly to the grading of the mathematics problems; there would be a unique solution manual for each unique exam and TAs would grade according to the logic and numeric values presented in the given exam.

\section{Case Study}\label{sec:results}

We ran a trial of our unique exam framework for the second midterm of our 10 week CS probability course at a R1 University. This course is a requirement for the undergraduate and master's CS degrees and often attracts 300 students, mostly at the sophomore-level in a CS or engineering-related field. In Spring 2020, our class consisted of 361 students. Our teaching team consisted of one instructor, and 13 teaching assistants. We had a public class forum for discussion and announcements, a course website for distributing course resources, and a Gradescope website~\cite{singh2017GradescopeFastFlexible} for managing submissions and grading.

In this iteration of the course, we gave students two 24-hour take-home online, exams --- one after week four, and one after week seven in our ten-week curriculum. Despite the large time window, our exam was designed for a more traditional, two-hour timed, in-person exam, with the understanding that students may need extended time due to timezone differences, technological barriers, or personal circumstance. Unique exams were only used for the second exam; our first exam was compromised within one hour of the 24-hour window.

Table \ref{tab:experience_timeline} shows a timeline and an estimated amount of time for each step in our process for deploying these unique and distinguishable exams to our students. This timeline can be used as a series of steps to implement the unique exams framework in one's own classroom. We close this case study with some lessons learned.

\begin{table}[tbp]
    \centering
    \begin{tabular}{|l|l|}
    \hline
    Task & Estimated time \\
    \hline
      1. Prepare exam problems & Twice the standard\\
      & preparation time \\
      2. Distribute distinguishable & Variable based on available \\
      \hspace*{2ex}exams & course infrastructure \\
      3. Grading & Comparable to standard \\
      &  grading time\\
      \hline
    \end{tabular}
    \caption{Timeline for deploying unique exam framework}
    \label{tab:experience_timeline}
\end{table}

\begin{figure}[h]
    \centering
    \includegraphics[width=3in]{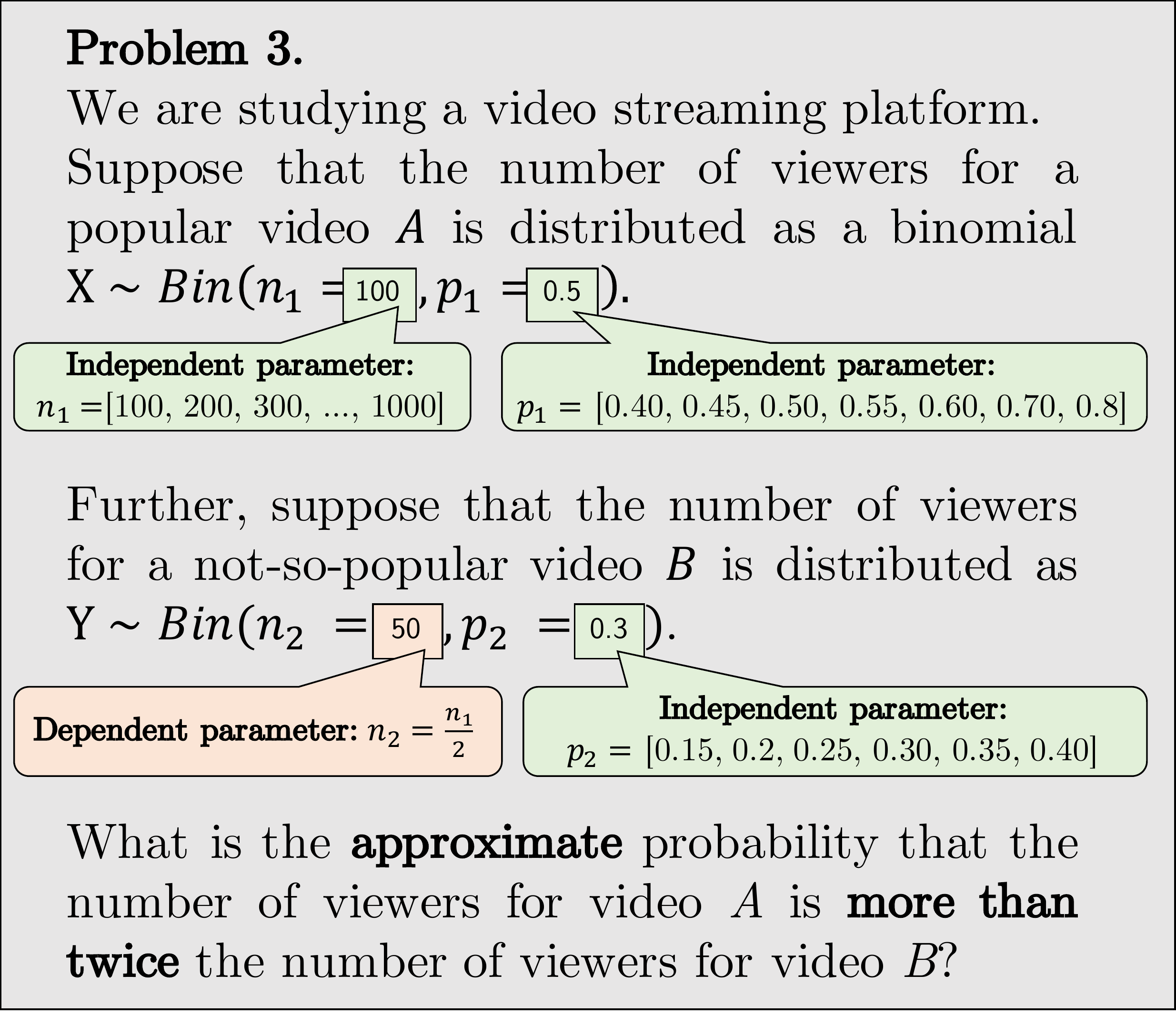}
    \includegraphics[width=3in]{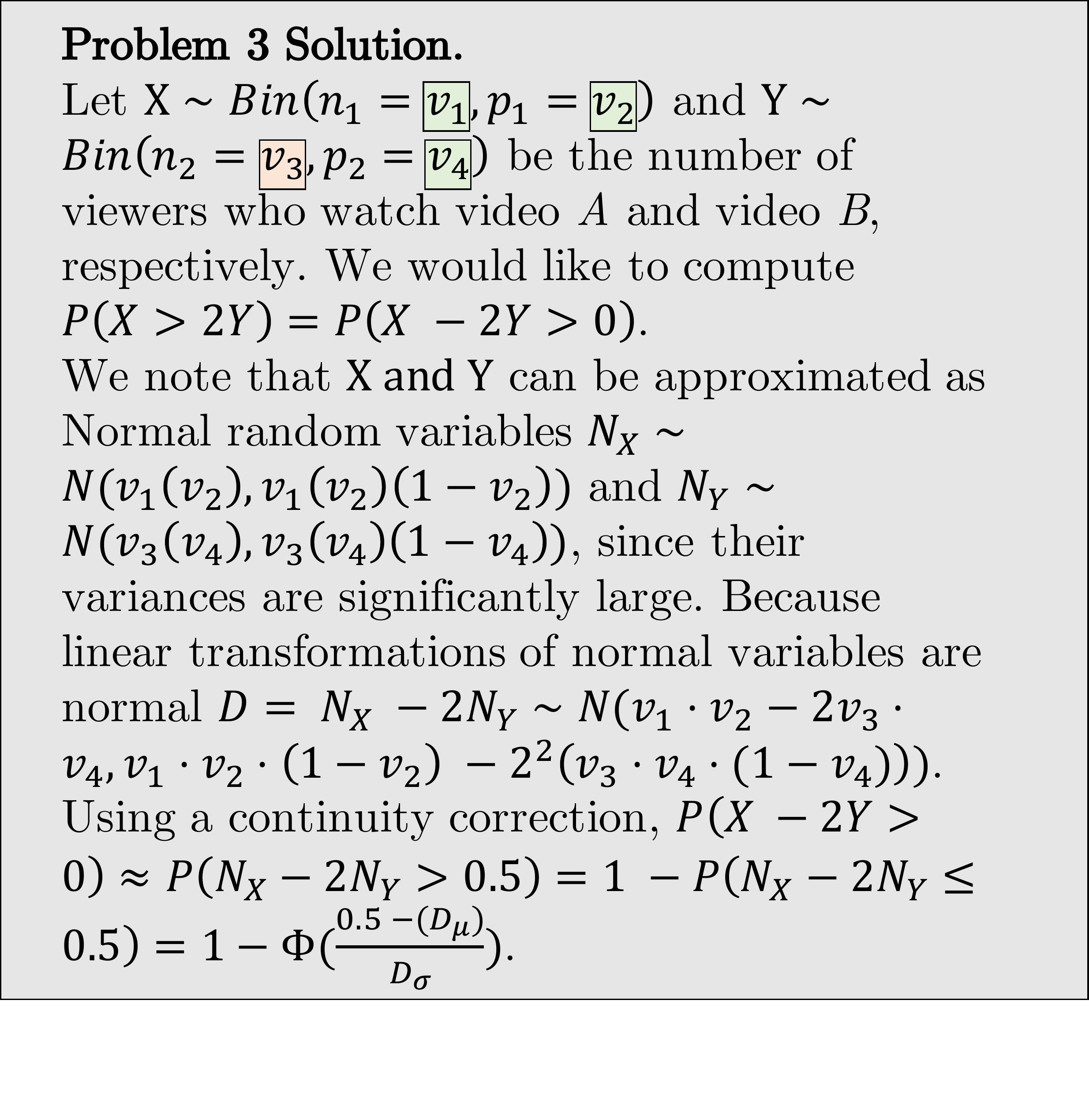}
   
        \begin{tabular}{|c|c|}
        \hline
          Parameter   & Type  \\
          \hline
            $n_1 = v_1 = [100, 200, ...,1000]$ & Independent Parameter \\
            $p_1 = v_2 = [0.40, 0.45, .. 0.55, 0.6, 0.7, 0.8]$ & Independent Parameter \\
              $n_2 = v_3 \frac{n_1}{2}$ & Dependent Parameter \\
                $p_2 = v_4 = [0.15, 0.2, ..., 0.4]$ & Independent Parameter \\
                  $n_1 (p_1 ) (1-p_1) > 10$ & Conditional Parameter \\
                   $n_2 (p_2 ) (1-p_2) > 10$ & Conditional Parameter \\
                    See diagram above & Solution \\
                  \hline
        \end{tabular}
        
    \caption{Problem skeleton, solution, and conditions for context dependent problem: binomial approximation example.}
    \label{fig:example_binom_problem}
\end{figure}
\subsection{Prepare exam problems: An end-to-end example}
Next, using our unique exam framework, we developed $361$ unique exams for the students in our class, where each exam had the same 4 problems (with about 4 subparts each). Each student received a unique exam that included different numeric values. The core problem was the same for each student.

Figure \ref{fig:example_binom_problem} shows the end-to-end implementation of a real binomial approximation problem in our exam, simplified for demonstration. This example highlights several interesting experiences in our exam design, and we refer back to it throughout this section.

In this example, the values of the random variables change the context of the problem. In class the students were taught that the Binomial distribution with parameters  $n$ and $p$ can be approximated by either the Poisson distribution ($n > 100, p < 0.05$) or the Gaussian distribution ($n > 20, np(1-p) > 10)$ depending on the parameters $n$ and $p$. Thus, in order to ensure that exams are of comparable difficulty, we need to ensure that the values we choose for $v_1$ and $v_2$ correspond to their intended meaning in the problem, independent of which numeric variations we use. This example is a simplified example with one step, but our actual exam problem was more complex, where the approximation was often a final step in a series of steps towards a solution. Therefore, as instructors, we must be careful with specific context that we cite in class and ensure that the experience among exams is uniform in difficulty and material tested.

Given the conditions detailed in Figure \ref{fig:example_binom_problem}, this problem generates $10 \cdot 7 \cdot 6 = 420$ unique problems. Since $21$ failed the conditional parameter, we had a total of $420 - 21 = 399$ valid problems. 

In our actual exam, for each of the four problems, we generated between 375 and 1200 unique problems. Since this was more than the number of students in our course, we selected 361 distinct versions of each problem resulting in 361 unique exams. 

In order to support our framework from exam release to grading and publishing, the instructor (she/her) had to first design the question as a word problem and write a solution. Next, she substituted in each numeric value (in both problem and solution) with a LaTeX variable representing an independent or dependent parameter. Finally, she provided a list of values for each independent parameter, formulas for generating dependent parameters, and rules for conditional parameters. The instructor and a subset of the course staff then iterated between modifying the problem statement and validating the unique problems created for logical consistency. Overall, this process took twice as long as the regular exam creation process due to solution-writing prior to exam release---which may be good practice regardless---and the iterative process of manually checking values.

\subsection{Distributing exam}
After developing $361$ unique exams, we distributed these unique and distinguishable exams online.

Prior to the 24-hour exam period, we sent students a stern message, stating that we expected students to solve all problems individually and specifically prohibited ``consultation of other humans in any form or medium (e.g., communicating with classmates, asking questions on forum websites...).'' ~\cite{corrigan2015deterring}. We also informed students that each of their exams will be unique and that if they violate the university honor code, they will be reported to the university's academic integrity oversight institution.

Students downloaded their unique version of the course exam through our course website, which was set up with university login credentials. Over the duration of the exam, we only allowed students to post questions and clarifications privately to the teaching staff on our course forum. Thus, we mitigated confusion that students might encounter from seeing each others' numbers referenced in question posts. We also kept an index of general clarifications about the exam which we posted publicly on our course forum.

\subsection{Grading exam}
Grading also proceeded as normal. We used our framework to generate a solution key spreadsheet with all unique intermediate steps towards the solution, mapped to each student. Grading for the quiz took approximately four hours, which is comparable to the grading time needed for the previous exam (3 hours), which was offered as a traditional exam.

After the exam was graded and published, students could use the same website application to download their uniquely and automatically generated solution key.

\subsection{Lessons learned}

The overall experience was positive and worked well for the students and course staff. As far as we can tell, there were no obvious forms of academic dishonesty---through online postings of the exam, which was the scenario we designed to deter---during this exam.

During the process of running these unique exams, we found that the binomial problem in Figure \ref{fig:example_binom_problem} had several unforeseen hurdles: certain numeric combinations for the parameters $n_1, p_1, p_2$ yielded values that were too large or too small to be computed using a standard calculator (e.g., Google calculator or WolframAlpha), and always yielded an answer of approximately 1.

\begin{figure}[tbp]
    \centering
    \includegraphics[width=0.7\columnwidth]{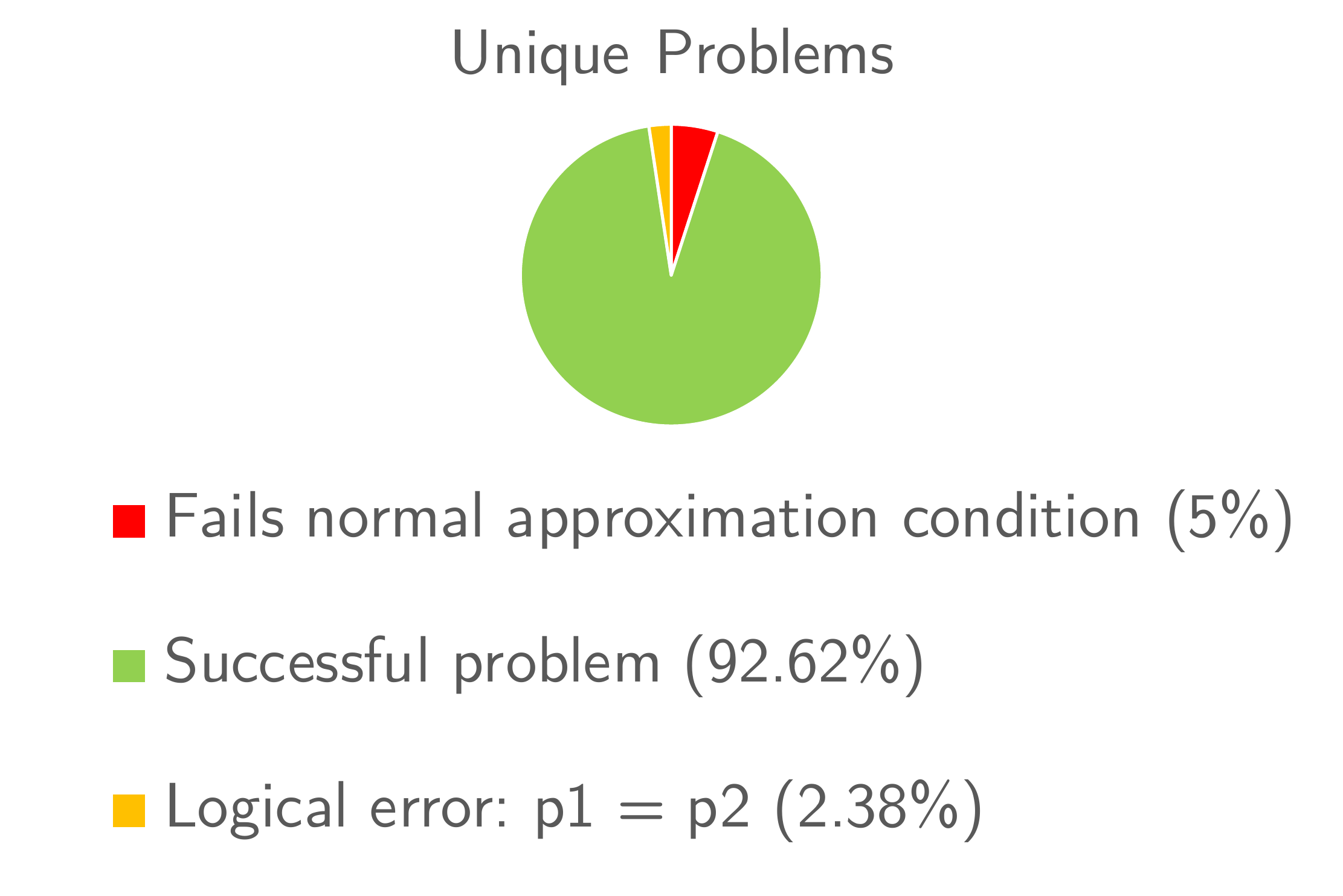}
    \caption{This figures summarizes the logical coherence of the 420 unique numeric combinations of the binomial approximation problem.}
    \label{fig:problem_difficulty}
\end{figure}

In addition, we found that some students were confused by the logic of the problem when $p_1 = p_2$ since in the context of the problem, this would mean that the more popular and less popular videos had equal probability of being watched. Figure \ref{fig:problem_difficulty} describes the relative feasibility of the different numeric combinations shown in Figure \ref{fig:example_binom_problem}.  In particular, of the $10 \cdot 7 \cdot 6 = 420$ different permutations of numeric combinations for the values of $n_1, p_1, p_2$, 21 combinations failed to satisfy the normal approximation condition where $np(1-p) > 10$. An additional 10 numeric combinations posed logical confusion to students where $p_1 = p_2 = 0.4$. However, most numeric combinations were both reasonable and logical.

Thus, we advise that instructors similarly check the problem feasibility from a calculation and logic points of view when using unique exams per student.

\section{Discussion}

Our system helps deter students from explicitly posting their problem on an open, online platform or discuss the problem among themselves. Since the numeric values in the problem have a uniquely identifiable tie to a particular student, students will know that explicitly posting the problem will lead to an identifiable academic integrity violations. Thus, the risk of getting caught outweighs the reward for possibly receiving the correct answer on one problem.

\subsection{Problem types}
The unique exams framework works best for problems that the instructor can write an algorithm to solve given the different constants in the problem. In particular, short-answer computational problems or code tracing problems can be easily and efficiently randomized, as shown in Section \ref{sec:method}. Computer science exams often also include long-form questions where students need to provide code segments or essay responses. For these problems, we argue that traditional plagiarism detectors or manual TA inspection by graders are more effective in securing these types of exams  \cite{yan2018tmoss}. Further, these types of problems might be better tested as take-home, open-resource projects with a compiler rather than timed exams. Evaluating the effectiveness of these different approaches to assessment is beyond the scope of this paper.

\subsection{Best practices}
% [Duration of exam] 
We gave students 24 hours to complete their exam, and we planned our exam to be the length of an in-person 2-hour exam. This allowed students the flexibility to work on the exam regardless of timezone, and to allow them to type their solutions without being concerned about running out of time.

% [Designing accurate numbers] 
An important consideration in using the unique exams framework is designing multiple distinct versions of the same problem skeleton that all have valid meaning with respect to the problem. Therefore, the instructor should carefully verify that each numeric value combination is appropriate for the problem.

% [Managing question / answer platforms during exams] 
It is important to allow students to ask the teaching team clarifying questions in an online exam. However, these questions should be made privately so that students are not confused by the numeric values present in their peers' exams.

% [Question bank] 
Instead of unique numeric values with the same exact question skeleton, other instructors use a question bank with distinct problems per student \cite{chen2020learning}. One advantage of a question bank is that it prevents excessive collaboration between students. However, unlike our solution, it allows for additional variability between student exams. Therefore, this might result in some exams being easier or harder than others. There are different tradeoffs. Depending on the particular type of exam administered, certain randomization methods might be more appropriate. 

\subsection{Limitations} We acknowledge that our current system does not prevent all forms of academic dishonesty, however. For example, a student could simply replace each numeric value in every problem with a variable name and post that question to an online platform or ask a friend.

It is possible that such behavior deters online community members or homework helpers from responding to the posting. 
One could also consider randomizing other parts of the given problem e.g. names within word problems, etc. This is akin to adding adversarial noise in developing adversarial examples for machine learning classifiers~\cite{goodfellow2015ExplainingHarnessingAdversarial}. 

Another solution is to try to make one of the values integral to the problem such that an outside helper would not have enough context without knowing that value. For example, in our course, we teach students two approximations of the Binomial distribution: for certain values of $n,p$, the Binomial distribution can be approximated as a Gaussian, while in other values of $n,p$ it is approximated by the Poisson distribution. Thus, if a student were to post a problem in an anonymous manner, an outside helper would not have enough context to solve this problem exactly.

Moreover, despite these possible approaches to deterring public, online posting of exams, the problem of collaborating \textit{offline} with classmates and experts remains. Our unique exam framework can be used in tandem with question bank-based exams, allowing for exam variation on multiple dimensions.

\subsection{Other solutions: rethinking the traditional exam model}
Different exam models work well for different courses. Traditional-form exams have been widely used in mathematics and CS theory courses, yet more implementation-based engineering courses may prefer other methods of assessing knowledge. We share our framework to port over the exam format to classrooms in which exams are \textit{ appropriate}, and other assessment models are prohibitively expensive to implement on a short timeline.

There are already ample examples of different frameworks that were successfully used at our university during our temporary remote instruction period. Our CS2 course exam was a one-on-one video programming interviews between teaching assistants and students because they supported a high teacher-student ratio. A discrete math course held individual pass/fail exams based on a revise-and-resubmit policy~\cite{stellmack2012ReviewReviseResubmit}. A third course in advanced systems had no exams, instead opting for grading only on programming assignment performance. Many more advanced courses in machine learning and systems opted for final projects in place of end-of-quarter exams.

\section{Conclusion}
We present the unique exams framework: a means to enable instructors to develop their own unique set of exams for their course. In just several simple steps, instructors can use our framework to help preserve integrity in their exams while preserving the positive student experience of timed feedback in a course. Without changing the process of the traditional exam, unique exams provide a shim layer of security for both students and instructors about exam reliability in the age of online education.

\bibliographystyle{ACM-Reference-Format}
\bibliography{biblio} 

\end{document}